\documentclass[]{raa}            
\usepackage{graphicx,times}
\usepackage{natbib}

\providecommand{\bm}[1]{\mbox{\boldmath$#1$\unboldmath}}

\def\etal{{et al.}}

\def\3o{O~{\sc ii}}                                               
\def\4o{O~{\sc iv}}

\begin{document}

   \title{Deriving the Coronal Hole Electron Temperature: Electron Density Dependent
Ionization/Recombination Considerations}

 \volnopage{ {\bf 2009} Vol.\ {\bf 9} No. {\bf XX}, 000--000}
   \setcounter{page}{1}

   \author{J.G. Doyle
      \inst{1}
   \and S. Chapman
      \inst{2}
   \and P. Bryans
      \inst{3,4}
   \and D. P\'erez-Su\'arez
      \inst{1}
   \and A. Singh
      \inst{1,5}
   \and  H. Summers
      \inst{6}      
   \and D.W. Savin
      \inst{7}     
   }

   \institute{Armagh Observatory, College Hill, Armagh BT61 9DG, N.
Ireland; {\it jgd@arm.ac.uk}\\
        \and
            Centre for Astrophysics, University of Central Lancashire,
Preston, UK 
        \and
            Space Science Division, Naval Research Laboratory,
Washington, DC 20375, USA
         \and
	 George Mason University, 4400 University Drive, Fairfax, VA 22020, USA
	 \and
	 Dept.of Physics \& Electronics, Deen Dayal Upadhyaya College, University of Delhi, India
	 \and
	 Department of Physics, University of Strathclyde, 107 Rottenrow, Glasgow, G4 0NG, Scotland
	 \and
	 Columbia Astrophysics Laboratory, 550 W 120th St, New York, NY 10027, USA
\vs \no
   {\small Received [year] [month] [day]; accepted [year] [month] [day] }
}

\abstract{Comparison of appropriate theoretical derived line ratios with observational 
data can yield estimates of a plasma's physical parameters, such as electron density 
or temperature. The usual practice in the calculation of the line ratio is the assumption 
of excitation by electrons/protons followed by radiative decay. Furthermore, it is normal 
to use the so-called coronal approximation, i.e. one only considers ionization and 
recombination to and from the ground-state. A more accurate treatment is to include the 
ionization/recombination to and from meta-stable levels. Here, we apply this to two 
lines from adjacent ionization stages; Mg~{\sc ix}~368~\AA\ and Mg~{\sc x}~625~\AA, which 
has been shown to be a very useful temperature diagnostic. At densities typical of coronal 
hole conditions, the difference between the electron temperature derived assuming the zero 
density limit compared with the electron density dependent ionization/recombination is  
small. This however is not the case for flares where the electron density is orders of 
magnitude larger. The derived temperature for the coronal hole at solar maximum is around 
1.04 MK compared to just below 0.82 MK at solar minimum.
}

   \authorrunning{J.G. Doyle et al. }            
   \titlerunning{A coronal temperature diagnostic}  
   \maketitle


%
%
\section{Introduction}           
\label{sect:intro}

Over the past few decades, many electron temperature and density diagnostics 
have been developed for application to observations of lines in the 
ultraviolet and extreme-ultraviolet regions of the spectrum, thereby providing 
insight to the physical parameters of the solar transition region and corona. 
The line ratio method usually involves a pair of lines from the same ion, 
although lines from adjacent ionization stages may be used. Thus, provided  
we have an accurate atomic model, the derived temperature provides essential 
data for an understanding of the basic processes occurring on the Sun.

Coronal holes, seen as regions with reduced emission due to a lower electron 
temperature than the surrounding quiet Sun, are an important source of 
open magnetic flux and high-speed solar wind plasma. Furthermore, observing the 
evolution of coronal holes over the course of a solar cycle can lead to a 
better understanding of the solar dynamo and advance our knowledge in 
forecasting space weather. However, all this requires an effective method to 
detect and track coronal hole boundaries. Harvey \& Recely (2002) used a 
hand-drawn method involving the He~{\sc i}~10830~\AA\ line plus photospheric 
magnetic field data. This was further developed by Henney \& Harvey (2005) 
and Malanushenko \& Jones (2005) showing good agreement between the hand 
drawn method and SoHO/EIT 195~\AA\ images. With the amount and quality of 
EUV data currently available from SoHO, a few attempts have been made to 
develop a more automated method of coronal hole identification.

Scholl \& Habbal (2008) used SoHO/EIT Fe~{\sc ix/x}~171~\AA, Fe~{\sc xii}~195~\AA\
and He~{\sc ii}~304~\AA, plus SoHO/MDI magnetograms. Chapman (2008, 2009) used
EUV spectral line data obtained with SoHO/CDS. He showed that the ratio of 
Mg~{\sc ix}~368~\AA\ and Mg~{\sc x}~625~\AA\ resonance lines are an excellent
temperature diagnostic for coronal holes, thus enabling a method
not only to locate the coronal hole boundary but also an evaluation of
the electron temperature with time. This work however employed the so-called 
coronal approximation, or zero-density approximation. This treats the populations 
of excited states of ions via an excitation balance of collisional excitation by 
electrons, and radiative decay. The ionization 
state is established as a balance of electron impact ionization from the ground 
state and radiative plus dielectronic recombination. In the simplest version of 
such modelling, secondary collisions with excited states are neglected.
\begin{figure*}[h!]                                            
\vspace{11.0cm}                                                
 \includegraphics{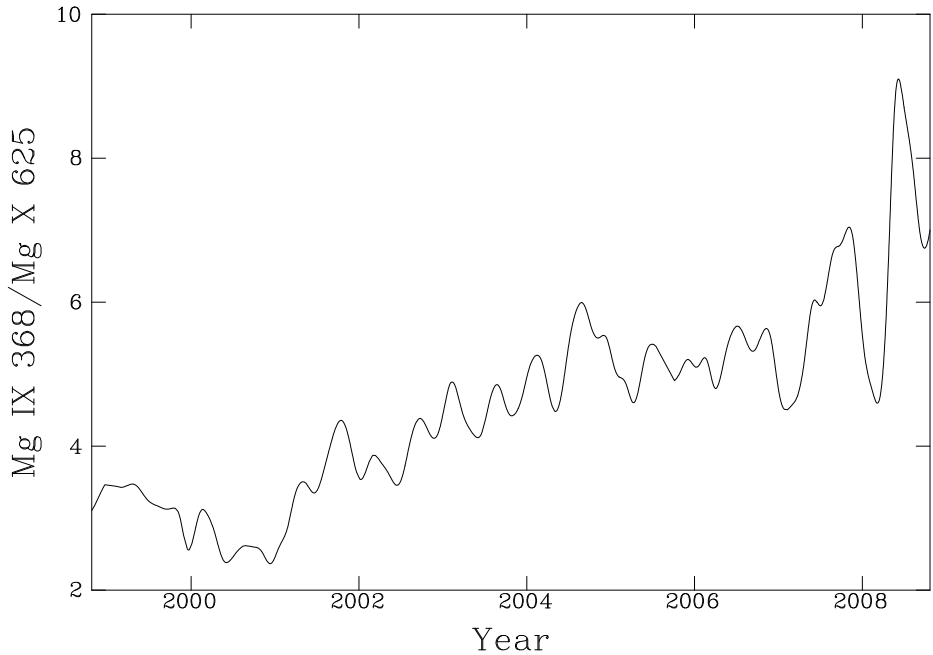}

\vspace*{-2.4cm}
\caption{The observed ratio of Mg~{\sc ix}~368~\AA\ to Mg~{\sc x}~625~\AA\ 
from 1998 onwards as derived by Chapman (2008, 2009) smoothed over 28 days.The typical error on this smoothed data is
less than 2\%. \label{Fig1}}
\end{figure*} 
Following on from work by Burgess \& Summers (1969) and Vernazza \& Raymond 
(1979), it was shown by Doyle \etal\ (2005, 2006) that the contribution functions 
of resonance lines from low Z Be-like and Li-like ions can be altered substantially 
due to electron density dependent ionization considerations. Although, we know 
that coronal holes are regions of reduced electron density compared to the 
quiet Sun, we do not know the extent of this additional atomic process with 
regard to Mg~{\sc ix} and Mg~{\sc x}. 

Here, we investigate this dependence for Mg and apply it to the synoptic 
data obtained with SoHO/CDS (Harrison \etal\ 1995) as discussed by Chapman (2008, 2009) in order 
to access the extent of the electron density sensitivity in the 
Mg~{\sc ix}~368~\AA\ and Mg~{\sc x}~625~\AA\ lines.


\section{Contribution Function}
\label{sect:CF}

An important component in calculating the line contribution function is 
the ionization balance.  As would be expected of analyses of the solar  
corona, the so-called coronal approximation is most often implemented when 
calculating the ionization balance of the emitting plasma.  In this        
approximation the plasma is considered optically-thin, low-density, dust-free,
and in steady-state or quasi-steady-state.  Under these conditions the        
effects of any radiation field can be ignored, three-body collisions are      
unimportant, and the ionization balance of the gas is time-independent.       
The population of ionization stages is calculated as a balance of ionization  
due to electron impact ionization (EII) and recombination due to radiative    
recombination (RR) and dielectronic recombination (DR). The most recent       
compilation of the ionization balance under these conditions was by Bryans    
\etal\ (2009).                                                                

However, under certain conditions, the low-density approximation may not be 
appropriate as activity in the solar atmosphere can result in an increase in 
density where the coronal approximation breaks down. In the coronal approximation 
there is no treatment of metastable states with                                   
populations comparable to the ground state.  In addition, the coronal             
approximation takes no account of secondary electron collisions. The              
generalized collisional-radiative picture (Summers \& Hooper, 1983) allows        
such an analysis. In detail, the collisional ionization and redistribution        
processes from excited states are included. The populated metastable states are   
determined via an elaborated ionization balance along with the ground states.     
These were computed within the Atomic Data and Analysis Structure (ADAS;          
Summers 2009\footnote{http://www.adas.ac.uk}) framework, which is a               
collection of fundamental and derived atomic data, and codes that manipulate      
them.                                                                             

Considering metastable level populations (including the ground level) to be
dynamic and excited levels (i.e. all other levels) to be quasi-static, one can form population equations
for each $N^{(z)}_{\rho}$ and $N^{(z)}_i$in terms of these matrix elements as                           
\begin{eqnarray}                                                                                        
\frac{dN^{(z)}_{\rho}}{dt} & = & C_{\rho\sigma}N^{(z)}_{\sigma} + C_{\rho                               
j}N^{(z)}_{j} + C_{\rho\sigma+}N^{(z+1)}_{\sigma+} +                                                    
C_{\rho\sigma-}N^{(z-1)}_{\sigma-} \label{crmatrix1}\\                                                  
0 & = & C_{i \sigma}N^{(z)}_{\sigma} + C_{i j}N^{(z)}_{j} + C_{i                                        
\sigma+}N^{(z+1)}_{\sigma+} + C_{i \sigma-}N^{(z-1)}_{\sigma-} \label{crmatrix2}                        
\end{eqnarray}                                                                                          

\noindent
where we denote metastable states (including the ground state) by Greek
indices and excited states by Roman indices. In the above we           
use the summation convention over repeated indices.                    
Such a matrix contains elements                                        
$C_{\rho \sigma}$, $C_{\rho j}$, $C_{i \sigma}$,                       
and $C_{ij}$ denoting the excitation                                   
(and de-excitation) rates between the respective levels;               
$C_{\rho \sigma_+}$ and $C_{i                                          
\sigma_+}$, denoting the sum of the recombination rates from each metastable;
and $C_{\rho \sigma_-}$ and $C_{i \sigma_-}$, denoting ionization rates;     
while the                                                                    
diagonal elements $C_{\rho \rho}$ and $C_{i i}$, indicate total
collisional and radiative loss rates
from the levels $\rho$ and $i$.

Recent work by Doyle
\etal\ (2005, 2006) looked at this problem regarding formation of the resonance
lines from low Z Li-like and Be-like ions. In the present work we use ADAS to
calculate the contribution functions of Mg~{\sc ix}~368~\AA\ and Mg~{\sc x}~625~\AA\ for
different values of the electron
density. The results of the ratio of these
contribution functions are shown in Fig.~2. We use the ionization and recombination data
from ADAS in these calculations. The variation of the contribution
functions is primarily due to the suppression of dielectronic recombination
with increasing density as collisional depopulation of high $n$-shells, formed
after stabilization, takes effect. This results in the ionization balance being
shifted to lower temperatures and is reflected in the contribution function.
In the next section, we apply the above results to recent coronal hole data.

\begin{figure*}[h!]                                            
\vspace{11.0cm}                                                
 \includegraphics{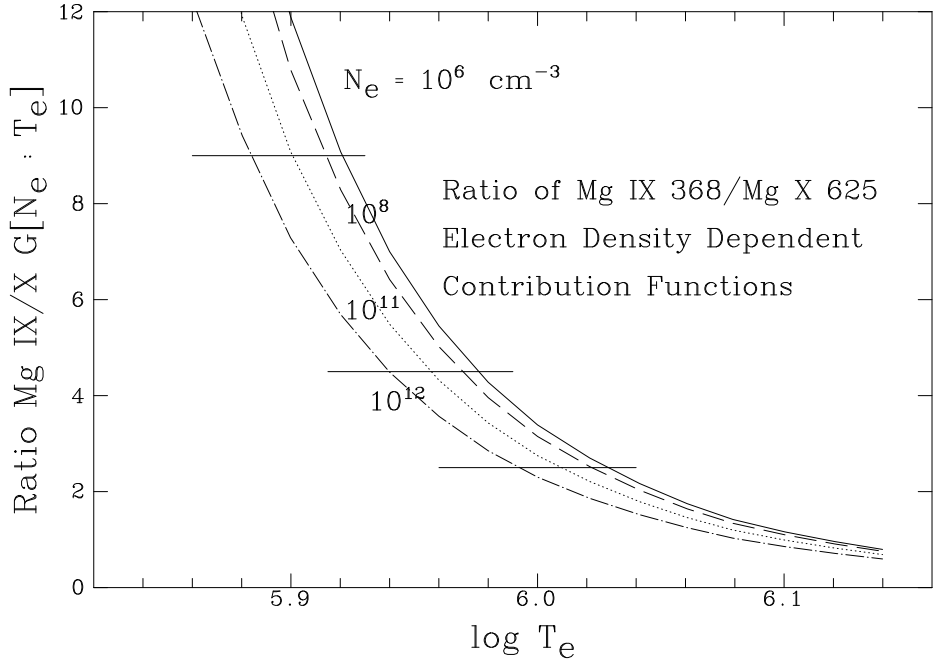}

\vspace*{-2.4cm}
\caption{The ratio of the electron density dependent contribution
function for Mg~{\sc ix}~368~\AA\ and Mg~{\sc x}~625~\AA\ for four
values of electron density ranging from very low densities to  
flare-type densities. The short horizontal lines at ratios of 1.0,
2.5 and 4.0 are the observed solar minimum, mean and maximum      
values, respectively (see Fig.~1).\label{Fig1}}                                
\end{figure*}

\begin{figure*}[h!]                                            
\vspace{11.0cm}                                                
 \includegraphics{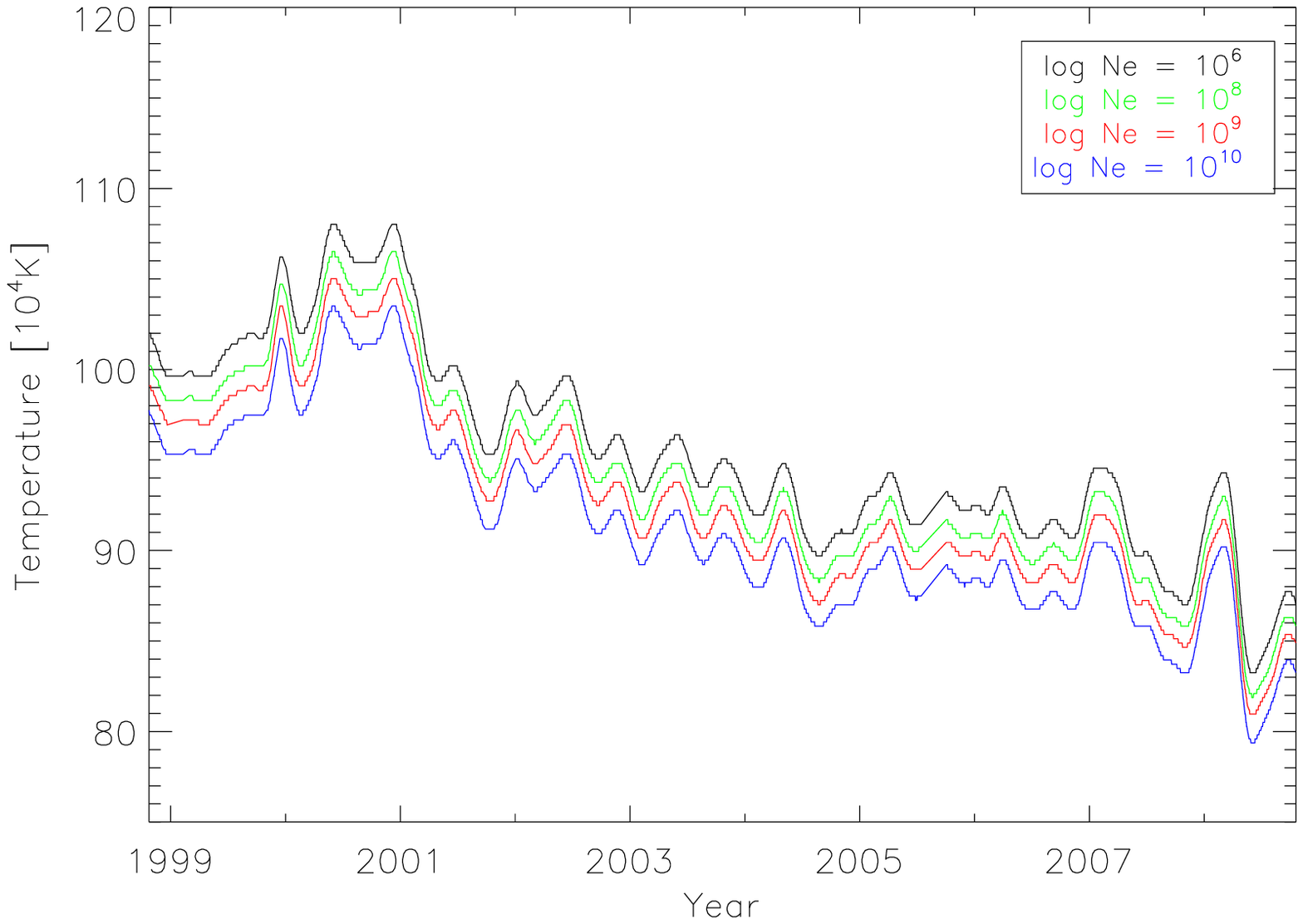}

\vspace*{-1.1cm}
\caption{The derived electron temperature over a 9 year period from
1998 onwards for different electron densities. Note the            
lower temperature at solar minimum ($\approx$ 2007/8), while at solar 
maximum ($\approx$ 2001/2) there are                                  
intervals where the electron temperature is above 1 MK.               
\label{Fig2}}                                                         
\end{figure*}       

\section{Observational Data}
\label{sect:data}

The observational data used by Chapman and reproduced here (see Fig.~1) were the
CDS synoptic study (SYNOP) which created 2 dimensional images and           
were run daily from around mid-night. The study generated a series          
of nine rasters each of which covers a 4$\times$4 arcmin area. Together, the        
nine rasters spanned the central meridian of the Sun. The study             
used a selection of the strongest lines available in the CDS/NIS            
wavelength range, covering a wide range of temperatures from the            
chromosphere to the corona: He~{\sc i}~584~\AA, O~{\sc v}~629~\AA, Mg~{\sc  
vii}~367~\AA, Mg~{\sc ix}~368~\AA, Mg~{\sc x}~625~\AA\ and  Fe~{\sc xvi}~361~\AA. As
a result, a near-continuous dataset exists since 1996 which is                      
well-suited to studying long-term variations in the solar structure.                
However, here we only use data taken after the SoHO recovery, i.e.                  
from 1998; due to cross-calibration problems which are discussed in more            
detail by Chapman (2008, 2009). Various methods used to extract the line            
fluxes, including background subtraction, use of small windows, line                
fitting, calibration, are all discussed by Chapman (2008, 2009).

\section{Results}
\label{sect:res}

At densities typical of coronal hole conditions (i.e. 10$^8$~cm$^{-3}$
to 10$^9$~cm$^{-3}$, see Raymond \& Doyle 1981), the difference of including ionization/recombination 
to/from meta-stable levels is small compared to the normal coronal approximation.
This however, is not the case for flares where the electron density can 
be orders of magnitude larger. In Fig. ~3, we show the derived electron 
temperature over a 9 year period from 1998 onwards for different electron 
densities. Note the lower temperature at solar minimum ($\approx$ 2008), while 
at solar maximum ($\approx$ 2001) there are intervals where the electron 
temperature is above 1 MK. Due to the high quality of the observational data, 
these observed changes, including not only the difference at solar minimum/maximum,
but also the short-term changes throughout the cycle are real. The paper by Chapman 
(2009) discusses in more detail the changes in the coronal hole coverage throughout 
the solar cycle, and its implications.

\vspace*{-0.2cm}
\normalem
\begin{acknowledgements}
Research at the Armagh Observatory is grant-aided by the N. Ireland Dept. of 
Culture, Arts and Leisure. We thank UK Science and Technology Facilities 
Council for support via
ST/F001843/1. SoHO is a mission of
international cooperation between ESA and NASA. PB was supported in part by
the NASA Solar and Heliospheric Physics Supporting Research and Technology program.
We thank A. Whiteford for his 
guidance on using ADAS. JGD thank the International Space Science Institute, 
Bern for the support of the team ``Small-scale transient phenomena  and their
contribution to coronal heating'
\end{acknowledgements}

\label{lastpage}

\end{document}